\documentclass[prl,aps,twocolumn,showpacs,floatfix,amsmath,amssymb,aps]{revtex4-2}

\usepackage{graphicx}% Include figure files
\usepackage{dcolumn}% Align table columns on decimal point
\usepackage{bm}% bold math
\usepackage[colorlinks]{hyperref}% add hypertext capabilities
\usepackage{multirow}
\usepackage{booktabs}
\usepackage{mathtools}

\hypersetup{colorlinks,linkcolor=blue,citecolor=blue,filecolor=blue,urlcolor=blue}
\newcommand{\ket}[1]{\left|#1\right\rangle}

\newcommand{\braket}[1]{\left\langle#1\right\rangle}
\newcommand{\bfS}{\boldsymbol{\sigma}}
\newcommand{\bfR}{\boldsymbol{R}}
\newcommand{\baone}{\boldsymbol{a}_1}
\newcommand{\batwo}{\boldsymbol{a}_2}
\newcommand{\bfb}{\boldsymbol{B}}

\newcommand{\bfk}{\boldsymbol{k}}
\newcommand{\chileft}{\chi^\triangleleft}
\newcommand{\chiright}{\chi^\triangleright}

\begin{document}
\title{Chiral Spin Liquid in Rydberg Atom Arrays}

\author{Yu-Feng Mao$^{1}$}
\author{Shicheng Ma$^{1}$}
\author{Yong Xu$^{1,2}$}
\email{yongxuphy@tsinghua.edu.cn}
\affiliation{$^{1}$Center for Quantum Information, IIIS, Tsinghua University, Beijing 100084, People's Republic of China}
\affiliation{$^{2}$Hefei National Laboratory, Hefei 230088, People's Republic of China}

\begin{abstract}
Despite long-standing theoretical interest, the chiral spin liquid, a topologically ordered phase,
has yet to be observed experimentally. Here we surprisingly find its emergence in 
an experimentally realized dipolar $\text{XY}$ model when Rydberg
atoms are arranged in a breathing kagome lattice.
Using the infinite density matrix renormalization group,
we numerically calculate the ground state's chiral order parameter, spin-spin correlations,
Chern number, and entanglement spectrum. 
Our numerical results provide strong evidence for the chiral spin liquid phase. 
Furthermore, we identify a quantum phase transition from a Dirac spin liquid to 
a chiral spin liquid as the lattice geometry is tuned from the isotropic kagome to the breathing 
kagome lattice.
\end{abstract}

\maketitle

Quantum spin liquids (QSLs) have garnered significant interest in condensed matter 
physics since their prediction~\cite{andersonResonating1973MRB,wenTOPOLOGICAL1990IJMPB,
balentsSpin2010N,savaryQuantum2016RPP,zhouQuantum2017RMP,knolleField2019ARCMP,broholmQuantumSpinLiquids2020,
kivelson502023NRP,lancasterQuantum2023CP}. 
They emerge from the melting of traditional magnetic 
order due to quantum fluctuations in frustrated spin systems at low temperatures~\cite{readLargeN1991PRL,sachdevKagome1992PRB}, eluding 
the conventional symmetry-breaking description~\cite{wittenQuantum1989CP,wenVacuum1989PRB,wenMeanfield1991PRB}. 
Various types of QSLs have been theoretically predicted. 
A notable example is the chiral spin liquid~\cite{kalmeyerEquivalence1987PRL,wenChiral1989PRB}, 
which displays a chiral order parameter 
by spontaneously breaking time-reversal symmetry (TRS) 
and exhibits ground-state degeneracy and 
semionic bulk excitations. Despite being proposed over three decades ago, 
the chiral spin liquid has not yet been experimentally observed.

Rydberg atom simulators have become a powerful platform~\cite{schaussQuantum2018QST,
saffmanQuantum2010RMPa,browaeysManybody2020NP} 
for simulating various spin models~\cite{barredoCoherent2015PRL,labuhnTunable2016N,
bernienProbing2017N,guardado-sanchezProbing2018PRX,lienhardObserving2018PRX,
de2019observation,schollQuantum2021N,semeghiniProbing2021S,chenContinuous2023N,gonzalez-cuadraObservation2025N,liangObservation2025PRL,everedProbing2025N,
qiaoRealization2025N,yueObservation2025,geimEngineering2026}.
Notably, the $\mathbb{Z}_2$ spin liquid phase has been prepared and observed in a 
Rydberg simulator~\cite{verresenPrediction2021PRX,semeghiniProbing2021S}. 
This raises the question of whether Rydberg atom arrays can also realize 
the chiral spin liquid state. 
It is well known that the $J_1$-$J_2$-$J_3$ XXZ antiferromagnetic model on a 
kagome lattice can support the chiral spin liquid phase~\cite{messioKagome2012PRL,gongEmergent2014SRa,
heChiral2014PRL,huVariational2015PRB,heDistinct2015PRL,gongGlobalPhaseDiagram2015,sunPossible2024nQM}. 
To address the issue of 
power-law-decaying dipolar interactions in Rydberg atoms,
a recent theoretical study~\cite{tianEngineering2025PRL} suggests using Floquet engineering to control 
the long-range interactions for model realization.
To reduce experimental complexity, it is worth considering whether the chiral 
spin liquid phase can emerge in the currently 
experimentally realized XY spin model with long-range dipolar interactions
without employing Floquet engineering. However, a recent study 
reveals that the model's ground state is the gapless $U(1)$ Dirac spin 
liquid rather than the chiral spin liquid~\cite{bintzDirac2024}. 
Very recently, this model has been experimentally realized in a Rydberg atom kagome array,
with observations revealing signatures beyond the ordered phase~\cite{bornetDirac2026}. 

\begin{figure}[ht]
	\includegraphics[width=\linewidth]{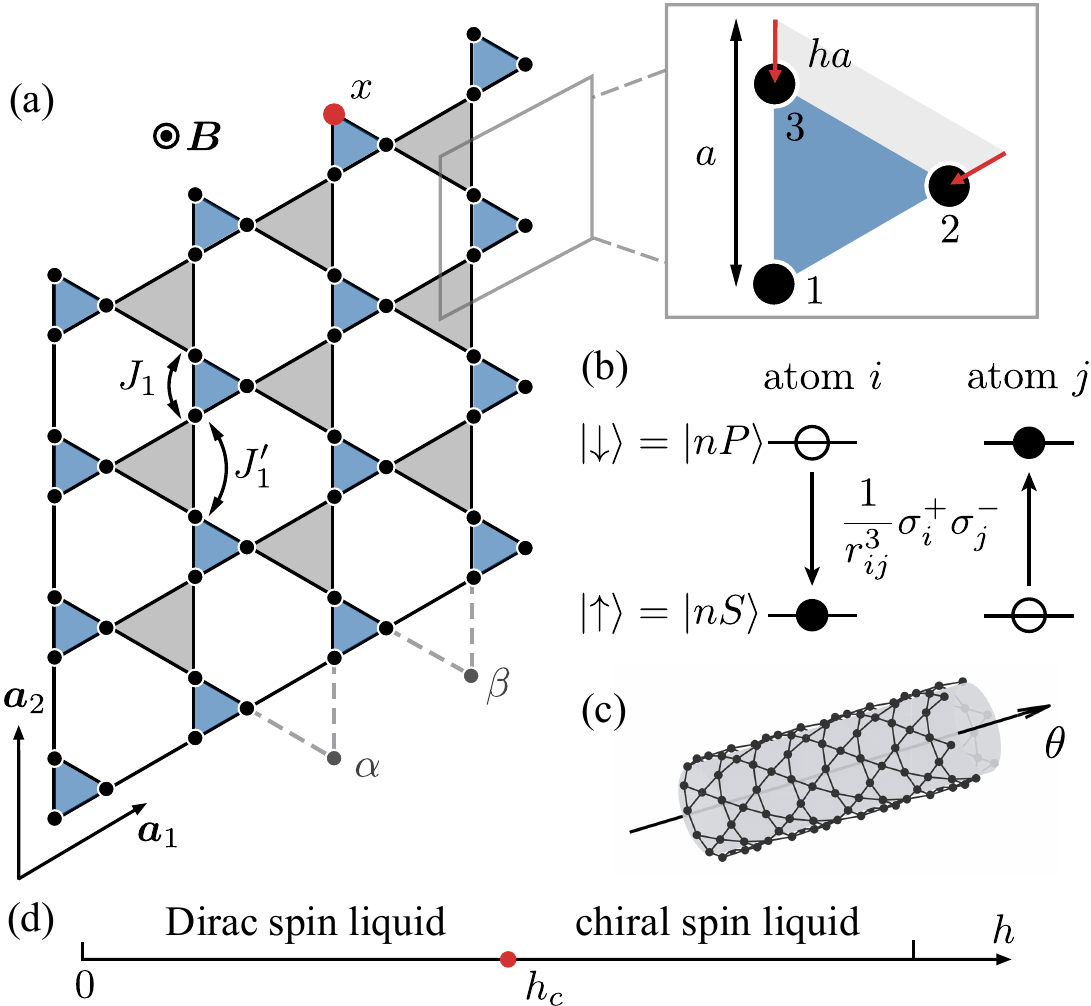}
	\caption{
		(a) Rydberg atoms are arranged in a breathing kagome lattice with
		a zoomed-in view of a unit cell.
		Within each unit cell, site 2 and site 3 are displaced from 
		their isotropic kagome lattice positions by $h a$, while site 1 remains fixed.
		Atomic displacements modulate the long-range dipolar interactions,
		e.g., splitting the nearest-neighbor coupling into
		two distinct strengths $ J_1 \sim 1/[(1-h)a]^3 $
		and  $ J_1^\prime \sim 1/[(1+h)a]^3 $.
		An external magnetic field $\bfb$ is applied perpendicular to the atomic plane.
		(b) Dipolar interactions between atom $i$ and atom $j$, 
		giving rise to an effective spin-$1/2$ XY model.
		(c) Cylindrical geometry employed in the iDMRG calculations.    
		Different cylindrical geometries, YC$2m$ and YC$2m$-2, 
		correspond to periodic boundary conditions along $\batwo$, obtained by 
		identifying site $x$ with site $\alpha$ or site $\beta$ 
		[see (a)], respectively. 
		A flux $\theta$ can be inserted by modifying the dipolar interaction terms to 
		probe the properties of distinct spin liquid phases.
		(d) Schematic phase diagram of the model in Eq.~(\ref{Hxy}) 
		with respect to the parameter $h$. 
	}\label{Fig1}
\end{figure}

Here, we surprisingly find that the ground state of
the dipolar XY model is a chiral spin liquid
when Rydberg atoms are arranged in a breathing kagome lattice~\cite{schafferQuantum2017PRB,repellinStability2017PRB,
iqbalPersistence2018PRB}, a slight deformation
of the kagome lattice [see Fig.~\ref{Fig1}(a)].
Our numerical calculations of the chiral order parameter, the spin-spin correlations,
the Chern number, and the entanglement spectrum using the infinite density 
matrix renormalization group (iDMRG)~\cite{whiteDensity1992PRL,ostlundThermodynamic1995PRLa,
mccullochInfinite2008,schollwockDensitymatrix2011AoP,
cincioCharacterizing2013PRL,grushinCharacterization2015PRB} provide strong evidence for 
the chiral spin liquid phase. In addition, we show that as the lattice geometry
changes from the kagome lattice to the breathing kagome lattice,
the ground state transitions from the Dirac spin liquid
to the chiral spin liquid state,
as evidenced by appearance of the chiral order parameter and a significant 
increase in the correlation length at the critical point.
Finally, we show that the state can be prepared by slowly reducing 
a staggered field in experiment settings.

\emph{Model Hamiltonian}---We start by considering a two-dimensional 
array of Rydberg atoms arranged in
a {\it breathing} kagome lattice, as shown in Fig.~\ref{Fig1}(a).
An effective spin-$1/2$ degree of freedom at each site is 
encoded in two Rydberg states,
$\ket{\uparrow}=\ket{n S}$ and $\ket{\downarrow}=\ket{n P}$,
where $n$ is a large principal quantum number.
The dominant interaction between atoms within this spin-$1/2$ 
manifold is the resonant dipole-dipole interaction, leading to
an effective dipolar XY model,
\begin{equation}
    {H}_{\mathrm{XY}}
     = \frac{1}{2}\sum_{i<j}{V_{ij}}\left(\sigma_i^x\sigma_j^x+\sigma_i^y\sigma_j^y\right),
    \label{Hxy}
\end{equation}
where $\sigma_i^{\nu}$ with $\nu=x,y,z$ are Pauli matrices at site $i$ [see Fig.~\ref{Fig1}(b)].
An external magnetic field $\bfb$, oriented perpendicular to the atomic array,
enforces isotropic dipolar interactions within the lattice plane.
The coupling is given by $V_{ij} = d^2 / R_{ij}^3$,
where $d$ is the dipole moment of the Rydberg states
and $R_{ij}$ is the distance between sites $i$ and $j$.
We take $d^{2}/a^{3}$, $\hbar a^3/d^2$, and $a$ to be the units of energy, 
time, and length, respectively, 
where $a$ is the lattice constant.
This Hamiltonian has been experimentally realized by using optical tweezers
to trap Rydberg atoms in a kagome lattice~\cite{bornetDirac2026}. 
The system conserves the total spin along $z$, $\sigma^z \equiv \sum_i \sigma_i^z$ or the
$U(1)$ symmetry, and 
we will henceforth restrict our analysis to the half-filling subspace where 
$\sigma^z = 0$. 

Breathing kagome lattices can be simply realized by adjusting the positions 
of optical tweezers in a Rydberg atom array, as illustrated in the inset of 
Fig.~\ref{Fig1}(a). Starting from an isotropic kagome lattice,
two atoms within each unit cell are displaced along
the lattice translation vectors $ \baone=(\sqrt{3}a,a)$ 
and $\batwo =(0,2a)$ by $-ha$, respectively, introducing breathing 
anisotropy. The system parameter $h$ thus controls
the degree of lattice deformation, 
which modifies the dipolar interactions between atoms (see Supplemental Material~\cite{supp})
and drives the phase transition shown in Fig.~\ref{Fig1}(d).

We investigate the ground-state properties of the Hamiltonian in the $\sigma^z=0$ subspace 
using iDMRG on lattices wrapped as an infinite cylinder 
with YC$2m$ or twisted YC$2m$-2 geometry~\cite{yanSpinLiquid2011Sa} [see Fig.~\ref{Fig1}(a,c)].
To capture the possible spontaneous TRS breaking,
we use complex-valued wave functions.
Long-range dipolar interactions are considered up to the seventh-nearest neighbors.
The bond dimension is chosen to be sufficiently large to ensure that
the maximal truncation error remains below $3\times10^{-5}$.

\begin{figure}[t]
    \includegraphics[width=\linewidth]{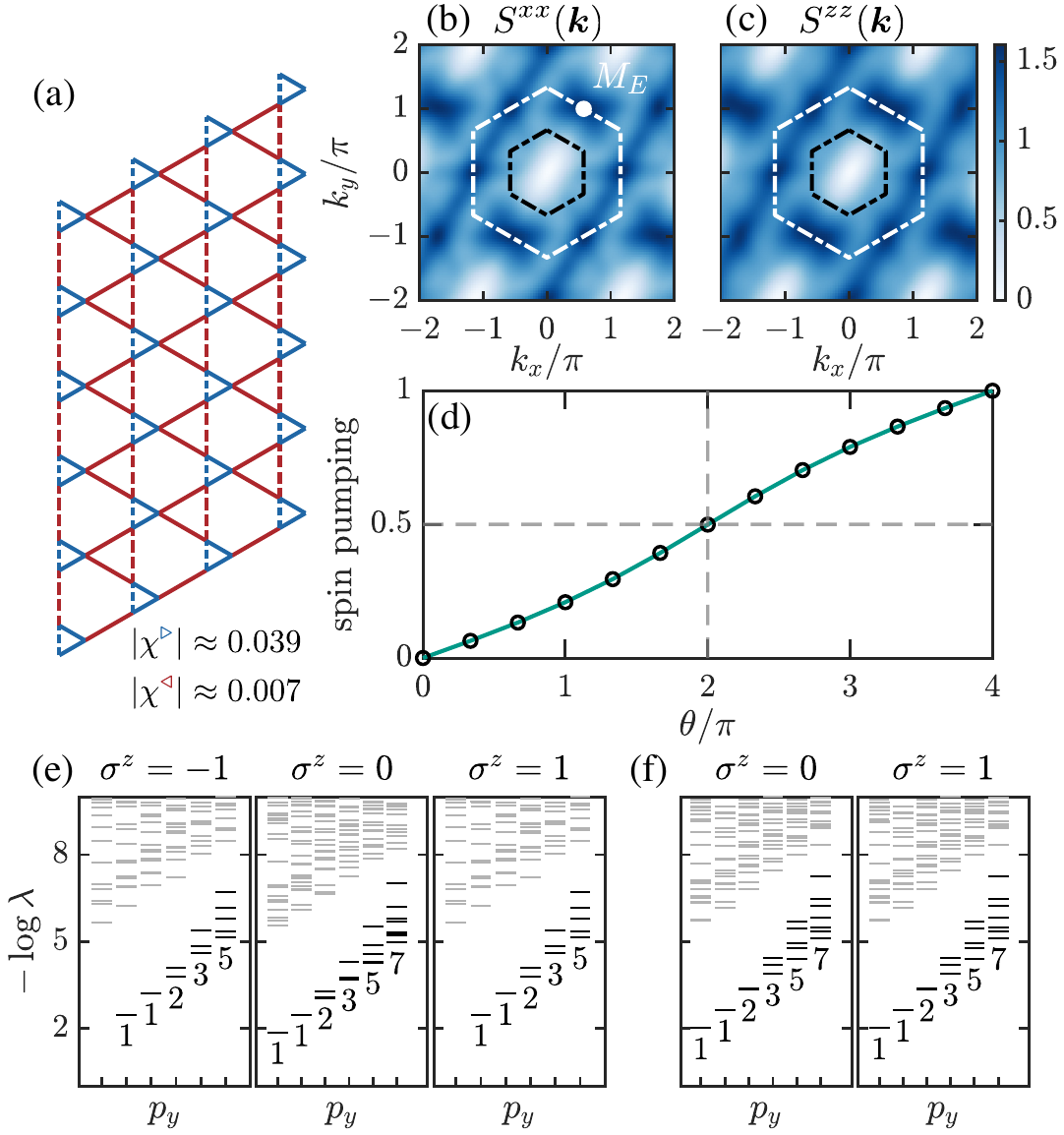}
    \caption{
        (a) The spin-spin correlations $\braket{\sigma^\nu_i \sigma^\nu_j}(\nu=x,z)$ plotted as 
        colored dashed or solid bonds between two neighbors with mean values
		(blue solid, blue dashed, red solid, red dashed):
		$\braket{\sigma_i^x \sigma_j^x} = (-0.336,-0.324,-0.204,-0.217)$,
		$\braket{\sigma_i^z \sigma_j^z} = (-0.332,-0.321,-0.177,-0.189)$.
        For separations beyond these neighbors, the magnitude of all correlations 
        is less than $0.06$. In addition, the chiral order parameters $\chiright$ 
        and $\chileft$ are provided.        
        Note that both states $\psi_1$ and $\psi_s$ yield qualitatively similar results.
        The spin structure factors (b) $S^{xx}(\bfk)$ and (c) $S^{zz}(\bfk)
        $ exhibit peaks at the $M_E$ point of the 
        extended Brillouin zone (white dashed line).
        The unit of $k_x$ and $k_y$ is $1/a$.
        (d) Spin pumping under adiabatically inserted flux $\theta$.
        Starting from one ground state, the flux $\theta$ is 
        increased in steps of $\pi/3$.
        The adiabaticity is verified by the fidelity 
        $\left|\braket{\psi(\theta)|\psi(\theta+\pi/3)\right|}|^2\approx 1$ 
        between adjacent flux steps. 
        Entanglement spectra of the YC12 cylinder for 
        (e) $\psi_1$ and (f) $\psi_s$, resolved by $\sigma^z$ and 
        $p_y = 0, 2\pi/6, \ldots, 5\times 2\pi/6$.
        The unit of $p_y$ is $1/(2a)$.
        The spectra display the chiral spin liquid characteristic counting 
        $\{1,1,2,3,5,7,\ldots\}$, marked in black.
        The spin-pumping results are obtained on the YC10 cylinder 
        with a bond dimension of $3000$, whereas all other results are extracted
         from the YC12 cylinder with a bond dimension of $10^{4}$. Here, we set $h=0.3$. 
    }\label{Fig2}
\end{figure}

\emph{Chiral spin liquid}---We now demonstrate that the system 
hosts a chiral spin liquid when $h=0.3$.
The chiral spin liquid is a gapped topological phase whose fractionalized quasiparticles 
are semionic spinons. On an infinite cylinder within our iDMRG calculations,
two topologically degenerate ground states are obtained, 
$\psi_1$ and $\psi_s$,
corresponding to the absence and presence of a spinon line 
threading the cylinder, respectively.

To establish the liquid nature of the phase, we calculate the
real-space spin-spin correlation functions
$\braket{\sigma_i^x \sigma_j^x}$ and $\braket{\sigma_i^z \sigma_j^z}$
of the ground state. We find that the correlations
decay significantly as the separation between the sites increases.
In addition, they exhibit
the same translation symmetry as the lattice [see Fig.~\ref{Fig2}(a)]. 
These results indicate the absence of the long-range magnetic order.
We also find a short correlation length $\xi \approx {1.6}a$ extracted from the iDMRG transfer matrix,
which agrees with the previous results for a chiral spin liquid~\cite{heChiral2014PRL,yaoQuantum2018NP}.
We further analyze the equal-time spin structure factor,
$ S^{xx}(\boldsymbol{k}) = (1/N) \sum_{i,j} 
e^{i \boldsymbol{k}\cdot(\bfR^0_i-\bfR^0_j)}\braket{\sigma_i^x \sigma_j^x}$,
where $\bfR^0_i$ denote the site coordinates of
the isotropic kagome lattice and $N$ is the number of sites, 
enabling direct comparison across different 
breathing kagome geometries.
Both $ S^{xx}(\boldsymbol{k}) $ and $S^{zz}(\boldsymbol{k})$
exhibit peaks near the edge-centered $M_E$ points of the extended Brillouin zone
[see Figs.~\ref{Fig2}(b,c)], which is consistent with the previous results~\cite{WZhu2018PNAS} 
and analogous to the results reported for the 
Dirac spin liquid~\cite{bintzDirac2024}.

A defining property of the chiral spin liquid is the spontaneous breaking of 
TRS, captured by the scalar chiral order parameter~\cite{wenChiral1989PRB}
\begin{equation}
    \chi_i^{\triangleright} = 
    \braket{\bfS_{i_1} \cdot (\bfS_{i_2}\times \bfS_{i_3})}/3,
\end{equation}
where $i_1$, $i_2$, and $i_3$ denote the three sites in the $i$th right 
triangle and similarly for $\chi_i^{\triangleleft}$.
The nonzero values of $\chi^{\triangleright}$ and $\chi^{\triangleleft}$ 
for both ground states are listed in Fig.~\ref{Fig2}(a), providing strong
evidence for the chiral spin liquid phase. Note that $|\chiright|$ is much 
larger than $|\chileft|$ due to the stronger interactions on the 
$\triangleright$ triangles.

We provide additional evidence for the chiral spin liquid phase by computing the Chern number.
Within the iDMRG framework, the Chern number is 
extracted from spin pumping, 
defined as the change in spin polarization 
across a fixed cut during a flux insertion~\cite{grushinCharacterization2015PRB}.
The flux is implemented by modifying the exchange interaction terms as 
$\sigma_i^+ \sigma_j^- \to e^{i\theta_{ij}}\sigma_i^+ \sigma_j^- $, 
where $\sigma_i^+$ and $\sigma_i^-$ are the spin raising and 
lowering operators, respectively, and 
$ \theta_{ij}= (\bfR_{ij}\cdot \bfR_\text{mod}/R_{\text{mod}}^2) \theta$, 
with $\bfR_\text{mod}$ being the periodic lattice vector around the cylinder~\cite{heSignatures2017PRX,bintzDirac2024}. 
Figure~\ref{Fig2}(d) shows that the pumped spin is quantized to $1/2$ and $1$
at $\theta=2\pi$ and $4\pi$, respectively,
establishing a fractional Chern number of $1/2$.
Notably, $\psi_1$ and $\psi_s$ are found to evolve into each other
during the $2\pi$ flux insertion~\cite{heObtaining2014PRB,heChiral2014PRL},
consistent with the properties of the $\nu=1/2$ fractional quantum Hall state.

Finally, we examine the entanglement spectrum, which provides 
a fingerprint of the chiral edge states. 
By performing a vertical cut on the infinite matrix product state (MPS), 
the reduced density matrix of a semi-infinite cylinder is obtained,
whose entanglement spectrum mimics the physical edge excitation spectrum~\cite{liEntanglement2008PRL,qiGeneral2012PRL}.
Owing to the conserved $U(1)$ symmetry and 
translation invariance along the cylinder axis $\boldsymbol{a}_2$, 
the spectrum is resolved in the good quantum numbers $\sigma^z$ and $p_y$~\cite{cincioCharacterizing2013PRL}.
As shown in Figs.~\ref{Fig2}(e-f), for each $\sigma^z$ sector,
the level counting at a fixed $p_y$ follows $\{1,1,2,3,5,7\ldots\}$,
consistent with the theoretical predictions for a chiral edge mode~\cite{mooreEdge1997PRB}.
Moreover, $\psi_1$ and $\psi_s$
exhibit distinct symmetry patterns in their entanglement spectrum:
$\psi_1$ is symmetric about $\sigma^z=0$, while $\psi_s$ exhibits identical 
entanglement spectrum for $\sigma^z=0$ and $\sigma^z=1$, reflecting the presence of a spinon line. 
Finally, the right- and left- moving spectra appear with equal weight 
in the iDMRG calculation, further confirming the spontaneous TRS breaking.

\begin{figure}[t]
    \includegraphics[width=\linewidth]{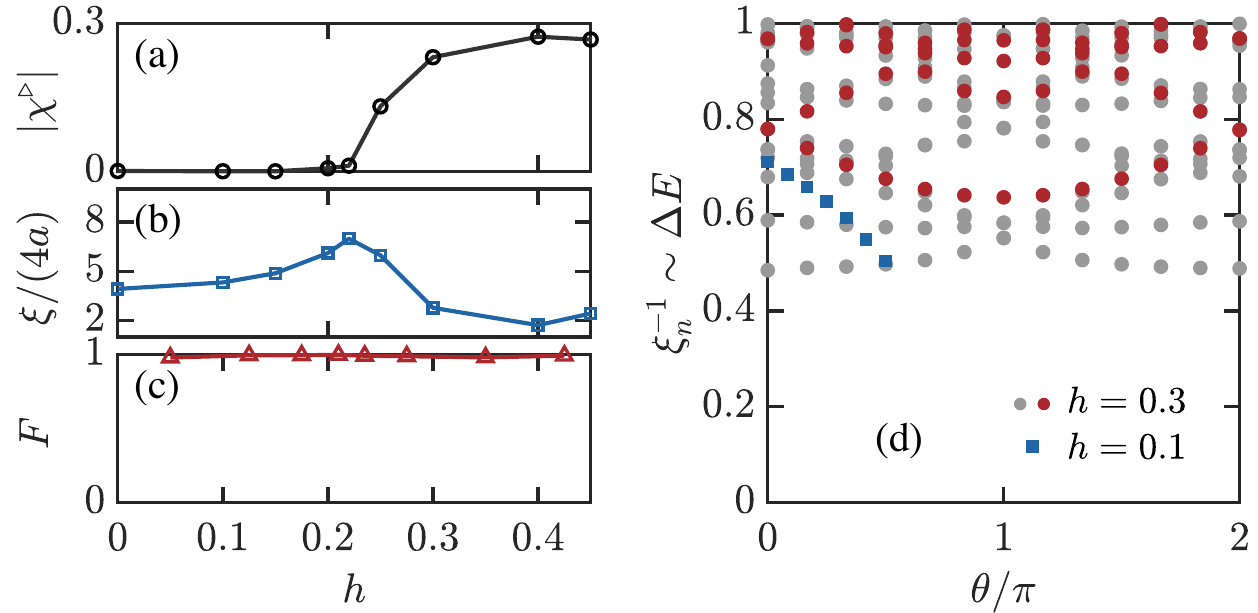}
    \caption{(a) The chiral order parameter,
    	(b) the correlation length, and
    	(c) the fidelity between neighboring points, with respect to the 
    	parameter $h$, revealing the continuous phase transition from the 
    	Dirac spin liquid to the chiral spin liquid.
    (d) The transfer matrix spectrum exhibits a gapped excitation energy 
    for the model at $h=0.3$ (gray and red circular markers).
    The red markers correspond to internode scattering.
    For comparison, we also plot the spectrum of internode scattering 
    represented by the blue square markers for 
    the lattice at $h=0.1$ in the Dirac spin liquid phase.
    Prior to the loss of adiabaticity at $\theta = \pi/2$, 
    a linear dispersion indicative of a gapless Dirac point is visible.
    The cylinder employed here is the YC8-2 lattice, 
    with a bond dimension of $5000$ for calculations of the phase diagram and 
    the spectrum at $h=0.3$, and with a bond dimension of $7000$ for calculating the 
    spectrum at $h=0.1$.
    }\label{Fig3}
\end{figure}

\emph{Transition between Dirac and chiral spin liquids}---We next 
examine the phase transition from the Dirac spin liquid,
reported in Ref.~\cite{bintzDirac2024}, to the chiral spin liquid, 
driven by tuning the breathing parameter $h$.
All results in this section are obtained on the YC8-2 cylinder geometry;
we note that different cylinder geometries may 
shift the critical point~\cite{heDistinct2015PRL} (see Supplemental Material~\cite{supp}).
To ensure that the states remain in the same topological sector,
the iDMRG calculations are initialized from a common MPS.

Since the chiral spin liquid breaks TRS spontaneously 
while the Dirac spin liquid~\cite{affleckLarge1988PRB,hastingsDirac2000PRB} preserves it,
the transition is naturally characterized 
by the chiral order parameter $|\chiright|$.
Figure~\ref{Fig3}(a) illustrates that when $h \gtrapprox 0.22$, 
$|\chiright|$ becomes nonzero,
signaling the spontaneous TRS breaking.
The correlation length exhibits a pronounced peak
at the critical point [Fig.~\ref{Fig3}(b)], consistent with a continuous phase transition.
Further evidence is provided by the fidelity between neighboring $h$ values,
which remains close to $1$ throughout, exhibiting no level crossing 
during the transition [Fig.~\ref{Fig3}(c)].

To further distinguish the two phases, we verify the gapped nature of the chiral spin liquid 
via the correlation length spectrum.
The eigenvalues of the iDMRG transfer matrix, $\tau_n=e^{i k_n-\xi_n^{-1}}$, 
encode the excitation energies $\Delta E \sim \xi_n^{-1}$ and their momenta $k_n$~\cite{zaunerTransfer2015NJP,heSignatures2017PRX}.
As $\theta$ scans from $0$ to $\pi$, sweeping through the entire Brillouin zone, 
the gap remains open throughout [Fig.~\ref{Fig3}(d)], 
in contrast to the linearly dispersing gapless excitations of the Dirac spin liquid~\cite{heSignatures2017PRX,zhuEntanglement2018SA,huDirac2019PRL,bintzDirac2024}.

\begin{figure}[t]
    \includegraphics[width=\linewidth]{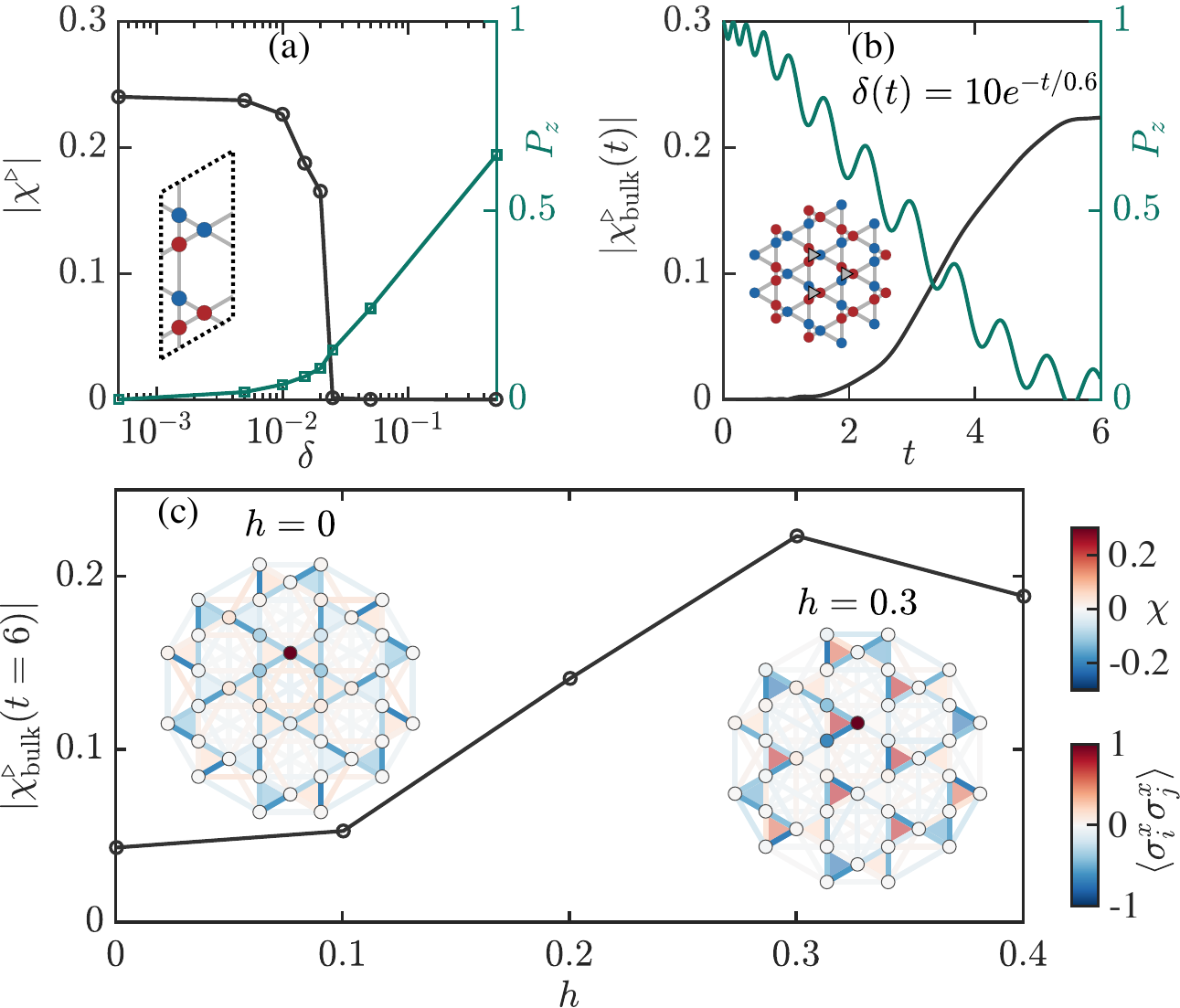}
    \caption{
    (a) Phase diagram of ${H}_\text{tot}$ on the YC8-2 cylinder at $h=0.3$.
    The inset shows the light-shift pattern assigned to each atom in the unit cell.
    (b) TDVP simulation of the quasi-adiabatic protocol 
    with $\delta_0 = 10$ and $\tau=0.6$ for a finite lattice of $N=42$ sites. 
    The inset shows the light-shift pattern applied to the lattice, 
    with blue (red) circles denoting $\eta_i=1 (-1)$.
    As the addressing light is ramped down, $P_z$ approaches zero and 
    a nonzero chiral order parameter in the bulk $\chiright_\text{bulk}$ emerges,
    where $\chiright_\text{bulk}$ is the average over the three central
    $\triangleright$ triangles marked gray in the inset.
    (c) Bulk chiral order parameter of the prepared state at $t=6$ as 
    a function of $h$.
    We see that $|\chiright_\text{bulk}|$ exhibits a significant rise as
    $h$ increases across the transition point.
    The left and right insets show the spatially resolved chiral order parameter 
    at each triangle and the correlations
    $\braket{\sigma_i^x \sigma_j^x}$ at $h=0$ and $h=0.3$, respectively.
    Site coordinates are fixed to those of the isotropic kagome lattice ($h = 0$) 
    to facilitate comparison across geometries.
    Each bond represents the correlation between connected sites; 
    circles indicate correlations relative to a fixed reference site highlighted by 
    the dark red circle.
	The time-evolved state is computed with a bond dimension of 1024.
    }\label{Fig4}
\end{figure}

\emph{Quasi-adiabatic preparation}---We now discuss how to prepare the chiral 
spin liquid state in experiments.
Analogous to the method employed for preparing the Dirac spin liquid~\cite{bintzDirac2024}, 
we consider the Hamiltonian
${H}_\text{tot}=H_\text{XY} + H^\prime$, where
$H^\prime= -\delta \sum_i \eta_i \sigma^z_i$ is a staggered perturbation 
with  $\eta_i\in\{-1,1\}$ being the staggered sign pattern [see the inset in Fig.~\ref{Fig4}(a-b)].
This perturbation can be realized via a spatially dependent AC Stark shift 
from focused pinning laser beams~\cite{chenContinuous2023N,qiaoRealization2025N,yueObservation2025,bornetDirac2026}.
Initially, we begin with an experimentally accessible product state 
within the $\sigma^z = 0$ subspace that approximates the ground state of 
${H}_\text{tot}$. 
The perturbation is then slowly ramped down to zero according to $\delta(t) = \delta_0 e^{-t/\tau}$, 
allowing the system to evolve towards the desired phase
while remaining near the instantaneous ground state.
Note that in experiments, the Hamiltonian $-{H}_\text{XY}$ is in fact 
realized, where the spin liquid phase appears in the highest energy state 
in the subspace $\sigma^z = 0$; however, this poses no challenge, as 
the initial state can be prepared in the highest energy state $H^\prime$,
as demonstrated in prior experiments~\cite{chenContinuous2023N,bornetDirac2026}.

The success of the protocol relies on the absence of any intervening phases 
during the ramp-down process.
To verify the absence of an intervening phase, we compute the phase diagram as a function of the field strength
 $\delta$, as shown in Fig.~\ref{Fig4}(a).
We see that the chiral spin liquid state emerges from an ordered phase 
as $\delta$ decreases, characterized by a sharp increase in the chiral 
order parameter and a significant decrease in 
$P_z = (1/N) \sum_i \eta_i \langle \sigma_i^z \rangle$. 
The existence of the transition is further corroborated by
a diverging correlation length at the critical point (see Supplemental Material~\cite{supp}). 
These findings confirm that no intermediate phase exists as $\delta$ changes.
Given that only a single phase transition separates the two phases, 
the ramp-down protocol provides a direct pathway to the target chiral spin liquid.

We employ the time-dependent variational principle (TDVP)~\cite{haegemanTimeDependent2011PRL,
haegemanUnifying2016PRB} method 
to numerically simulate the preparation protocol 
for a finite system of $N = 42$ sites [see inset of Fig.~\ref{Fig4}(b)].
As the external staggered field is gradually reduced, 
$P_z$ decays to zero while the chiral order parameter $\chiright$ in the bulk grows.
The final energy of the prepared state satisfies $E \approx 0.94 E_0$; 
the small deviation from the exact ground state energy ($E_0$) is attributed to 
the partial loss of adiabaticity near the critical point.
The final state is shown in the right inset of Fig.~\ref{Fig4}(c), 
displaying the spatially resolved chiral order parameter and 
spin-spin correlations up to third-neighbor distance.
The state exhibits strong chiral order with short-ranged correlations, 
consistent with the properties of the chiral spin liquid.
For comparison, the left inset of Fig.~\ref{Fig4}(c) shows 
the prepared Dirac spin liquid state, 
which exhibits a near-zero chiral order parameter in the bulk. 
We note that both states display a staggered chiral order parameter 
at the edges, which may be attributed to boundary effects.
Figure~\ref{Fig4}(c) further displays the mean bulk chiral order parameter at $t=6$
with respect to the parameter $h$, showing the emergence of chiral order across
the transition from the Dirac spin liquid to the chiral spin liquid.

A recent experiment has remarkably realized the Hamiltonian in Eq.~(\ref{Hxy})
using an {\it isotropic} kagome array with dipolar Rydberg atoms~\cite{bornetDirac2026}.
By reconfiguring the atoms into a breathing kagome geometry, the chiral spin liquid 
phase becomes accessible with essentially the same state-preparation procedure. 
Spin-spin correlations can be used to probe the liquid nature of this chiral spin 
liquid phase. Furthermore, measuring the chiral order parameter provides a 
direct signature that clearly distinguishes the chiral spin liquid from the Dirac spin liquid.

In summary, we have predicted the existence of a chiral spin liquid phase 
in a Rydberg atom array. Strong numerical evidence supports the existence of this 
phase and the phase transition between it and the previously identified Dirac 
spin liquid phase. Additionally, we demonstrate that the chiral spin liquid state 
can be experimentally realized on current state-of-the-art Rydberg atom 
quantum simulators. Apart from Rydberg atoms, such a Hamiltonian may also 
be realized in other quantum simulators, including polar molecules~\cite{mosesNew2017NP,
baoDipolar2023S,hollandOndemand2023S,cornishQuantum2024NP} and trapped ions~\cite{monroeProgrammable2021RMP,
fengContinuous2023N,guoSiteresolved2024N}.

\begin{acknowledgments}
We thank Z. Yue, M. K. Tey, K. Li, and N. Y. Yao for helpful discussions. 
This work is supported by the National Natural Science Foundation of China (Grant No. 12474265) and
the Innovation Program for Quantum Science and Technology (Grant No. 2021ZD0301604).
Calculations were performed using the TeNPy Library~\cite{tenpy2024}.
We also acknowledge the support by Center of High Performance Computing, 
Tsinghua University.
\end{acknowledgments}

\newpage
\begin{widetext}
	%%%%%%%%%% Prefix a "S" to all equations, figures, tables and reset the counter %%%%%%%%%%
	\setcounter{equation}{0} \setcounter{figure}{0} \setcounter{table}{0} %
	\renewcommand{\theequation}{S\arabic{equation}}
	\renewcommand{\thefigure}{S%
		\arabic{figure}} \renewcommand{\bibnumfmt}[1]{[S#1]}
	\renewcommand{\thetable}{S%
		\arabic{table}} \renewcommand{\bibnumfmt}[1]{[S#1]}
	\renewcommand{\thesection}{S%
		\arabic{section}} \renewcommand{\bibnumfmt}[1]{[S#1]}
	%\renewcommand{\citenumfont}[1]{S#1}
	%%%%%%%%%% Prefix a "S" to all equations, figures, tables and reset the counter %%%%%%%%%%
	% \section{Supplementary Material}
	
	\newpage
	In the Supplemental Material, we provide 
	the variation of interaction strengths induced by the lattice deformation in Section S-1 
	and iDMRG details and additional results in Section S-2.
	
	\section{S-1. Variation of interaction strengths induced by the lattice deformation}\label{S1}
	
	In this section, we provide more details about the breathing kagome lattice 
	and the variation of interaction strengths induced by the lattice deformation. 
	
	\begin{figure}[h]
		\includegraphics[width=\linewidth]{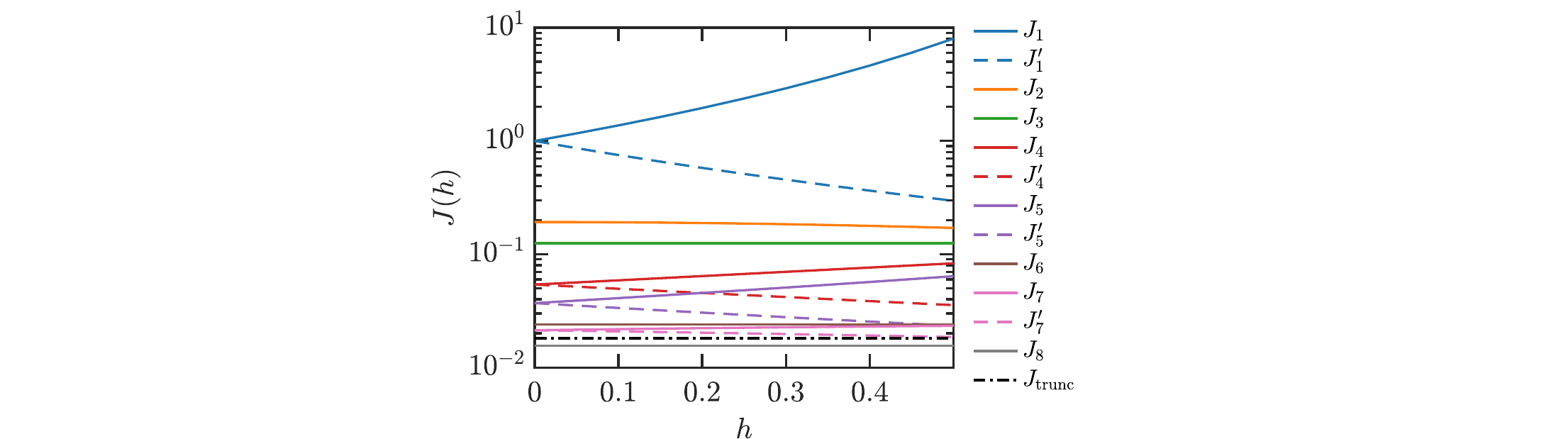}
		\caption{
			The strengths of the dipolar interactions up to the eighth-neighbor are shown as 
			a function of the parameter $h$. 
			The $n$th largest interaction of the isotropic kagome lattice may split into 
			two distinct values in the breathing kagome lattice: 
			a stronger interaction $J_n$ (solid line) and 
			a weaker interaction $J_n^\prime$ (dashed line). 
			The dot-dashed line indicates the truncation threshold 
			employed in the iDMRG calculations.
		}\label{FigS-geo}
	\end{figure}
	
	Figure 1(a) in the main text illustrates a breathing kagome lattice with
	three lattice sites in each unit cell labeled by sublattice indices $\{s = 1, 2, 3\}$. 
	The Bravais lattice vectors are chosen as 
	$\baone = a({\mathbf{e}}_x + \sqrt{3}\,{\mathbf{e}}_y)$ 
	and $\batwo = 2a\,{\mathbf{e}}_y$, where ${\mathbf{e}}_x$ and ${\mathbf{e}}_y$
	are unit vectors along $x$ and $y$, respectively. 
	Introducing the deformation parameter $h$, 
	the positions of the three sites within a unit cell are given by:
	\begin{equation}
		\begin{aligned}
			\bfR(i_x,i_y,1) & =  i_x \baone+i_y\batwo  \\
			\bfR(i_x,i_y,2) & =  (i_x+\frac{1}{2}-\frac{h}{2}) \baone + i_y \batwo   \\
			\bfR(i_x,i_y,3) & =  i_x \baone + (i_y + \frac{1}{2}-\frac{h}{2})\batwo, \\
		\end{aligned}
	\end{equation}
	where $i_x,i_y\in \mathbb{Z}$.
	
	When $h=0$, the isotropic kagome lattice respects the $D_6$ symmetry and the translational 
	symmetry. For $h\neq 0$, the translational symmetry is still preserved, while the $D_6$
	symmetry is reduced to the $D_3$ symmetry. Apart from the spatial symmetries, 
	the Hamiltonian on the lattice also respects
	the time-reversal symmetry (TRS) and the $U(1) \rtimes \mathbb{Z}_2$ symmetry, where 
	$U(1)={e^{i\theta \sigma^z}}$ and $\mathbb{Z}_2=\{1,\prod_j \sigma_j^x \}$.
	
	The structural deformation of the lattice gives rise to 
	a variation in the interaction strengths, 
	owing to the inverse-cube dipolar interaction between Rydberg atoms.
	In Fig.~\ref{FigS-geo}(b), we illustrate how the interaction strengths 
	vary with the breathing kagome lattice parameter $h$.
	For example, the nearest-neighbor interactions of the isotropic kagome lattice 
	are split into two distinct values: $J_1 \sim 1/[(1-h)a]^3$ 
	and $J_1^{\prime} \sim 1/[(1+h)a]^3$, 
	corresponding to the intra-unit-cell and inter-unit-cell 
	nearest-neighbor interaction magnitudes, respectively. 
	Moreover, as the parameter $h$ increases, 
	the ordering of long-range interactions can be altered: 
	Specifically, the weaker component $J_4^\prime(h)$ may fall below $J_5(h)$ 
	when $h>0.2$. 
	Here, $J_n$ denotes the $n$th largest interaction strength 
	in the isotropic kagome lattice, 
	which may split into a stronger component $J_n(h)$ and 
	a weaker component $J_n^{\prime}(h)$ in the breathing kagome lattice. 
	
	To ensure consistency in the number of interactions retained 
	across iDMRG calculations at different values of $h$, 
	we adopt a truncation scheme that includes all interactions 
	between atoms separated by a distance of at most 
	$R_{\text{trunc}} = 3.81\,a$. 
	Under this scheme, interactions up to $J_7$ are always retained 
	within the range of lattice deformation ($h$)  considered, 
	while those of order $J_8$ and beyond are neglected. 
	For TDVP calculations on finite lattices, 
	all long-range interactions are included without truncation.

	\section{S-2. iDMRG details and additional results}\label{S3}
	In this section, we discuss the details of the iDMRG calculations 
	used in the main text 
	and provide additional analysis of the 
	phase transition between the Dirac spin liquid and the chiral spin liquid. 
	We also present the divergence of the correlation 
	length at the critical point between the chiral spin liquid and the ordered phase, 
	which is relevant to the state preparation protocol 
	discussed in the main text.
	
	\subsection{A. iDMRG setup and convergence}
	\begin{figure}[h]
		\includegraphics[width=\linewidth]{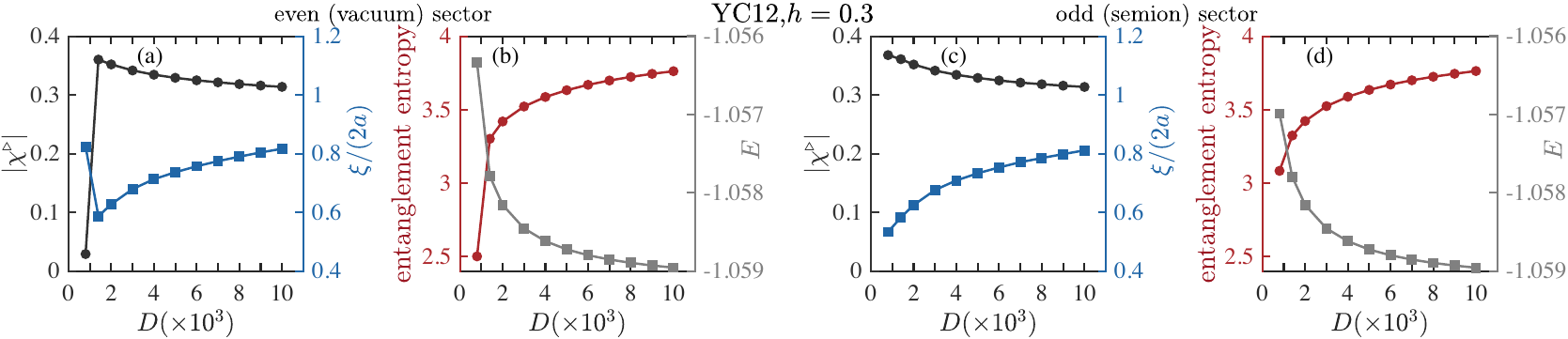}
		\caption{ 
			Convergence of the chiral order parameter ($\chiright$),
			correlation length ($\xi$), entanglement entropy, 
			and energy ($E$) as a function of a bond dimension $D$, 
			for the YC12 cylinder at $h = 0.3$. 
			Results are shown for (a,b) the vacuum sector  
			and (c,d) the semion sector.
		}\label{FigS-Convergence}
	\end{figure}
	Before presenting the details of the iDMRG calculations, 
	we note the approximations inherent in the numerical method. 
	The true ground state resides in an infinite system with all 
	long-range interactions retained; 
	in practice one must contend with a finite cylinder width, 
	a truncated interaction range, and a finite bond dimension. 
	To obtain reliable results, we include interactions up to 
	a considerable range, and the chiral spin liquid phase 
	is studied on cylinders up to YC12 with 
	a bond dimension of $10^4$ to ensure convergence 
	(see Fig.~\ref{FigS-Convergence}).
	
	To avoid trapping in local minima, we initialize 
	the calculations from a large number of random initial states, 
	first evolving each to a bond dimension of $2000$. We find that 
	all these random states converge to an identical state. 
	We then use this state as an initial state to perform calculations
	for larger bond dimensions. 
	In the two-site iDMRG calculations, a mixer is applied. 
	We also employ Dirac spin liquid wavefunctions as alternative initial states
	in the chiral spin liquid regime
	and find that the converged state is a chiral spin liquid state. 
	
    It has been reported that the ground states of the Dirac spin liquid 
	on an infinite cylinder also exhibit topological degeneracy, 
	residing either in the even or odd sector~\cite{huCompetingSpinLiquid2015,
		zhuSpin2015PRB,saadatmandSymmetry2016PRB,gongGlobal2017PRB,huDirac2019PRL}. 
	The entanglement spectra of these two sectors are symmetric 
	about $\sigma^z = 0$ and $\sigma^z = 0, 1$, respectively. 
	We find that initializing the calculation in the even sector 
	consistently yields the chiral spin liquid state in the vacuum sector, 
	whereas initialization in the odd sector yields the state in 
	the semion sector. 
	
	\subsection{B. Finite-size effects on the Dirac spin liquid to chiral spin liquid phase transition }
	\begin{figure}[h]
		\includegraphics[width=\linewidth]{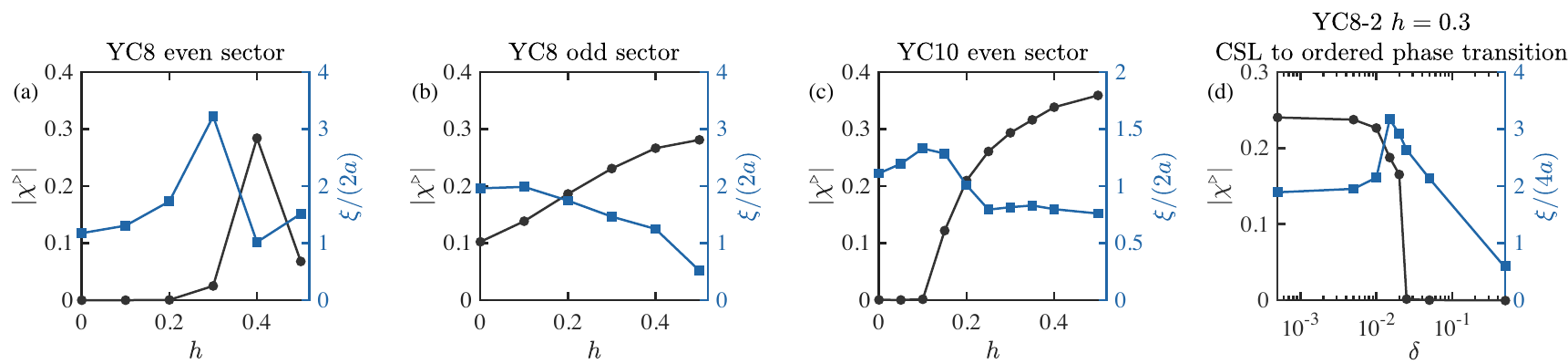}
		\caption{ 
			Phase diagram of the Dirac spin liquid to chiral spin liquid transition calculated 
			on (a) the YC8 even sector, 
			(b) YC8 odd sector, and (c) YC10 even sector. 
			In (a) and (c), the transition from the Dirac spin liquid to the chiral spin liquid is 
			signaled by the emergence of the chiral order parameter ($\chiright$),
			accompanied by 
			a significant increase in the correlation length ($\xi$) at the critical point. 
			In the YC8 odd sector (b), no phase transition is observed. 
			(d) The correlation length and the chiral order parameter as a function of 
			the perturbation strength $\delta$ for the Hamiltonian $H_{\text{tot}}$ in the main text.
		}\label{FigS-PT}
	\end{figure}

	We now present the phase diagrams calculated on the YC8 and YC10 cylinders in 
	Figs.~\ref{FigS-PT}(a--c).
	On YC8-2 and YC10, the odd and even sector states are exactly 
	degenerate, as they are related by translational symmetry~\cite{zaletelConstraints2015PRL}; 
	we therefore show only the even sector for YC10. 
	In the YC8 odd sector, the Dirac spin liquid has a higher energy than the chiral spin liquid at small $h$, 
	so no Dirac spin liquid to chiral spin liquid transition is observed in this sector [see Fig.~\ref{FigS-PT}(b)].
	For wider cylinders, YC10 and YC12, the Dirac spin liquid phase is recovered at $h = 0$ 
	in both sectors. We also see that compared to the result on YC8-2, the transition point 
	decreases to approximately $0.1$, indicating large finite-size effects on the precise transition point.
	Note that computing the phase diagram on YC12 is highly 
	computationally expensive and is therefore not included here.
	
	\subsection{C. The correlation length in the adiabatic preparation process}
	
	In the adiabatic preparation process, we apply a light-shift perturbation $H^\prime(\delta)$.
	Here, we provide the correlation length as a function of 
	the perturbation strength in Fig.~\ref{FigS-PT}(d). 
	The correlation length exhibits a single divergence along the entire chiral spin liquid preparation path, 
	with the peak coinciding with the vanishing of the chiral order parameter, 
	consistent with the existence of only a single phase transition 
	separating the two phases discussed in the main text.
	
\end{widetext}
\end{document}